\documentclass[
reprint,
bibnotes,
amsmath,amssymb,
aps,
prl,
]{revtex4-1}

\pdfoutput=1 

\usepackage{graphicx}
\usepackage{dcolumn}
\usepackage{bm}
\usepackage{xcolor}

\begin{document}

\title{Symmetry Breaking in Pedestrian Dynamics}

\author{Nickolas A. Morton}
\email{nmor487@aucklanduni.ac.nz}
\author{Shaun C. Hendy}
\email{shaun.hendy@auckland.ac.nz}
\affiliation{%
Te P\={u}naha Matatini, Department of Physics, University of Auckland, New Zealand\\
}%




\date{\today}

\begin{abstract}
When two pedestrians travelling in opposite directions approach one another, each must decide on which side (the left or the right) they will attempt to pass. If both make the same choice then passing can be completed with ease, while if they make opposite choices an embarrassing stand-off or collision can occur. Pedestrians who encounter each other frequently can establish ``social norms" that bias this decision. In this study we investigate the effect of binary decision-making by pedestrians when passing on the dynamics of pedestrian flows in order to study the emergence of a social norm in crowds with a mixture of individual biases. Such a situation may arise, for instance, when individuals from different communities mix at a large sporting event or at transport hubs. We construct a phase diagram that shows that a social norm can still emerge provided pedestrians are sufficiently attentive to the choices of others in the crowd. We show that this collective behaviour has the potential to greatly influence the dynamics of pedestrians, including the breaking of symmetry by the formation of lanes. 

\end{abstract}

\maketitle


When two pedestrians travelling in opposite directions approach one another, each must decide on which side (the left or the right) they will attempt to pass. If both make the same choice then passing can be completed with ease, while if they make opposite choices an embarrassing stand-off or collision can occur. Observations of pedestrians in different parts of the world have shown that communities tend to establish ``social norms" \cite{moussaid2009experimental, zanlungo2012microscopic} that bias the decision as pedestrians learn to anticipate the preferences of others. However, it is also known that different norms can be established in different communities \cite{zanlungo2012microscopic}: citizens of Osaka are reported to have the opposite bias to residents of Tokyo, something that is associated with the side of the escalator that pedestrians prefer to stand \cite{Downes2012,Ryall2015}. Simulations of pedestrian flows in the presence of such a bias have demonstrated that it can influence crowd behaviour and may be linked to the spontaneous formation of lanes of co-moving pedestrians \cite{helbing2002simulation}. 

Here we investigate the effect of binary decision-making \cite{Brock2007} by pedestrians when passing on the dynamics of pedestrian flows in order to study the emergence of a social norm in crowds with a mixture of individual biases. Such a situation might occur, for instance, when individuals from different communities mix at a large sporting event or at transport hubs such as railway stations or airports. We construct a phase diagram showing that a social norm can still emerge provided pedestrians are sufficiently attentive to the choices of others in the crowd, potentially having a significant influence on pedestrian dynamics. 

Spontaneous self-organisation is a feature of both biological \cite{Kondo1616} and social systems \cite{hemelrijk2005self}. When groups of pedestrians walking in opposing directions meet, they could self-organise into single unidirectional lanes to minimise the chance of collisions \cite{helbing2002simulation}, but this has the disadvantage that a pedestrian may find themselves behind a slower walker or in front of a faster impatient walker. Instead, if pedestrians on a busy street segregate entirely into two columns moving in opposite directions, they have an opportunity to pass without meeting others coming in the opposite direction. In order to do so they must break symmetry by establishing a collective preference for the left or the right.  

Modelling pedestrian flows has become vital for the design of public spaces \cite{zheng2009modeling}, and is also likely to become increasingly important as autonomous vehicles become more prevalent \cite{chen2016predictive}. There are a variety of models for pedestrian dynamics, including both macroscopic models that attempt to model pedestrian flows as a continuum (see \cite{xia2009dynamic} for instance) and microscopic models \cite{cristiani2014multiscale} that simulate individual pedestrian behaviour. We employ the so-called social physics model \cite{helbing1995social, castellano2009statistical}, which is a microscopic description of pedestrians who interact via forces. We believe the model presented here is the first microscopic model to feature decision-making for passing in pedestrian behaviour. While other models for pedestrians exhibit the formation of lanes (e.g. \cite{0256-307X-28-10-108901}), this behaviour has not been linked to the emergence of a preference for passing. Some studies consider crowds where established biases for passing were inherent \cite{zanlungo2012microscopic}, having originated as the result of prior social interactions. In our model, we consider crowds with a distribution of individual biases, but where each individual anticipates and may take into account the preferences of others \cite{Brock2007}.  

We begin by considering the encounter between two pedestrians, $i$ and $j$, who approach each other from opposite directions. To pass their counterpart, each pedestrian must choose to move to the left or the right. If they make the same choice, the passing manoeuvre will proceed smoothly; if they make opposite choices they risk an embarrassing collision. For a pedestrian $i$, we assume that the the difference in utility between choosing to pass pedestrian $j$ on the left or the right is given by: 
\begin{equation}
\label{ising}
\Delta U_i = U_i (L) - U_i (R) = h_i + J_i m_{ij} - \beta_i^{-1} \epsilon_i
\end{equation}
where $h_i$ is some externally imposed preference (e.g. a sign directing people to walk on the left), $m_{ij}$ is the decision of the opposing pedestrian $j$ as expected by $i$ (and $J_i>0$ is a measure of the embarrassment of a wrong choice), and $\epsilon_i$ is the pedestrians own preference for the left or the right (where $\beta>0$ is a measure of the attentiveness of the pedestrian). The preference $\epsilon_i \in [-1,1]$ where $+1$ is a strong preference for passing on the left. If $\Delta U > 0$, that is if $\epsilon_i < \beta \left(h_i + J_i m_{ij} \right)$, the pedestrian $i$ will try to pass $j$ on the left, and vice versa. 

Now assume that in a group of pedestrians $j \in N$ the preferences $\epsilon_j$ follow some distribution $f_\epsilon$. In this case, the probability that a pedestrian $i$ will try to pass on the left is $P_i\left(L\right) = F_\epsilon\left(\beta\left(h_i + J_i m_{ij} \right)\right)$, where $F_\epsilon$ is the cumulative distribution for $\epsilon$. That is:
\begin{equation}
P_i\left(L\right) =  \int^{\beta\left(h_i + J_i m_{ij} \right)}_{-1}{f_\epsilon d\epsilon} 
\end{equation}
if $\beta | h_i + J_i m_{ij} | < 1$, $P_i\left(L\right) = 1$ if $\beta\left(h_i + J_i m_{ij} \right) > 1$ or $P_i\left(L\right) = 0$ if $\beta\left(h_i + J_i m_{ij} \right) < -1$. If the oncoming pedestrian is also drawn from this group, then the expected preference of this pedestrian will in turn then be $m_{ij}^* = 2F_\epsilon\left(\beta\left(h_j + J_j m_{ji} \right)\right).$

Pedestrian $i$ is not necessarily aware of the expected preference of the opposing pedestrian, so in general $m_{ij} \neq m_{ij}^*$.  However, over time, if a group of pedestrians repeatedly encounter each other they may learn the preferences of others. In this case, pedestrians $i$ and $j$ may learn to expect that $m_{ij}=m_{ij}^*$ and $m_{ji}=m_{ji}^*$ respectively. If $h_i=h$ and $J_i=J$ are uniform across the population we then obtain an implicit equation for $m_{ij}$ \cite{Brock2007}:  
\begin{equation}
m_{ij} = 2F_\epsilon \left(\beta \left( h + J m_{ij} \right) \right).
\label{mean-field}
\end{equation}

We consider two simple examples. In the first, we assume that pedestrian preferences $\epsilon_{i}$ are uniformly distributed between $1$ and $-1$. In this case, (\ref{mean-field}) reduces to $m^\star = \beta \left(h + J m^\star \right)$, where $m^\star=m_{ij}$, or explicitly $m^\star = \beta h/\left( 1- \beta J \right)$. In this case, we see that pedestrian expectations are proportional to the external influence $h$ so if $h=0$ then pedestrian expectations are also $0$. In this example there is no spontaneous symmetry breaking, although symmetry breaking will occur in response to external influences $h \neq 0$.

If pedestrian preferences are polarised then more interesting behaviour can occur. Consider the distribution $f_{\epsilon} = \cosh \epsilon / \left(\sinh 1 \right)$, which gives an implicit equation for $m^\star$:
\begin{equation}
m^\star = \frac{\sinh \beta \left(h + J m^\star \right)}{\sinh 1}.
\label{weiss}
\end{equation}
For $h = 0$, there is a critical value of $\beta J$ above which symmetry breaking will occur spontaneously: namely, if $\beta J > \sinh 1$ then there are two solutions $m^\star_+ = -m^\star_- \neq 0$ to equation (\ref{weiss}), while if $\beta J < \sinh 1$ the only solution is $m^\star = 0$. This is the equivalent of the Curie point in the Curie-Weiss mean field theory for ferromagnetism. The quantity $\beta J$ can be thought of as the relative attentiveness of pedestrians to other's preferences: if the crowd is attentive ($J \gg \beta^{-1}$) then symmetry breaking will occur and the crowd will develop a social norm for passing the left or the right, while if the crowd is inattentive ($J \ll \beta^{-1}$ ) then no social norm will emerge.    

We now investigate the influence of such a decision model in simulations of pedestrian dynamics. We use a simple social force model \cite{helbing1995social}, in which we assume that the pedestrians are point charges who travel in a certain direction with a preferred walking speed but are repelled by one another. Each pedestrian has a direction of motion $n_{i} \in [1,-1]$ (fixed for the duration of the simulation), and a preferred passing direction $\omega_{i}$ $\in$ $[1,-1]$ (which can change). Passing is simulated via two sight cones, one with an effective radius of 2m for passing oncoming pedestrians, and a second with an effective radius of 1m for passing slower pedestrians moving in the same direction. Each pedestrian feels an additional force due to others within their sight cone, which acts either to the left or the right across the corridor depending on their passing preference. 

The corresponding equations of motion were integrated using the Verlet method with a time step of 0.01s. Pedestrian velocities were rescaled at each time step to ensure that pedestrians move at their preferred velocity. The initial conditions of each simulation were set by distributing 60 pedestrians within a long narrow corridor (20m by 2m, with periodic boundary conditions in the long axis) in two columns. Each pedestrian was allocated a preferred walking speed chosen from a Gaussian distribution with mean $\bar{\bf{v}}$ and standard deviation $\sigma_p$, with half wishing to travel north and the other half wishing to travel south. Observations of crowds at the University of Sydney found $|\bar{\bf{v}}|=1.34$ ms$^{-1}$ with a standard deviation of $\sigma = 0.26$ ms$^{-1}$ \cite{henderson1971statistics}, similar to the experimental values found in Ref.~\cite{moussaid2012traffic}. We use this value of $|\bar{\bf{v}}|$ while varying $\sigma_p$. A pedestrian's intrinsic preference for passing, $\epsilon_{i}$, is selected from a distribution $f_{\epsilon} = \cosh \epsilon / \left(\sinh 1 \right)$. 

A pedestrian's current passing direction, $\omega_{i}(t) =\mbox{sign}(h+ J m_{i}^\star (t) - \beta^{-1} \epsilon_i)$, is calculated using the binary decision model (\ref{ising}) at an interval $\Delta t =$0.5s (or 50 time steps) during a simulation. The timescale, $\Delta t$, was chosen to be similar to the relaxation time for a pedestrian coming to a halt to avoid a potential collision \cite{moussaid2011simple}. The expected pedestrian passing preference $m_{i}$ is calculated by averaging over the preferences of other pedestrians in line of sight of the pedestrian $i$ (defined as a cone in front of the pedestrian of 180 degrees arc and a radius of 2m). The preferences of pedestrians who are closer is weighted more highly than those who are more distant. We also include a delay, so that past encounters do have an influence on the expectation: $m_{i}^\star (t) = \left( 2/3 \right) \Delta m_i + \left(1/3 \right) m_{i}^\star(t-\Delta t)$ where the sum
\begin{equation}
\nonumber
\Delta m_{i} (t)  = ({\sum_{j=1}^n \dfrac{1}{r_{ij}^2}})^{-1} \sum_{j=1}^{n} \dfrac{1}{r_{ij}^2}\omega_{j}(t-\Delta t)
\end{equation}
runs over the $n$ pedestrians within the line of sight of pedestrian $i$. We think of this model as simulating an observant pedestrian, who is taking into account the passing behaviour of others nearby. To describe the ordering of the pedestrian flows in the corridor, we define an order parameter as follows: 
\begin{equation}
\label{OrderParameter}
O_t = \frac{1}{2N^2}\left( (\sum_{x_{i} < h} n_{i})^2 + (\sum_{x_{i} > h} n_{i})^2 \right)
\end{equation}    
where $h$ is the half width of the corridor and $N$ is the total number of pedestrians in the simulation. We also monitor the mean of the individual {\em expected} preference $\langle m_{i}^{*} \rangle$ as an indication of the emergence of a social norm in the corridor.

\begin{figure}[h]
\resizebox{\columnwidth}{!}{\includegraphics{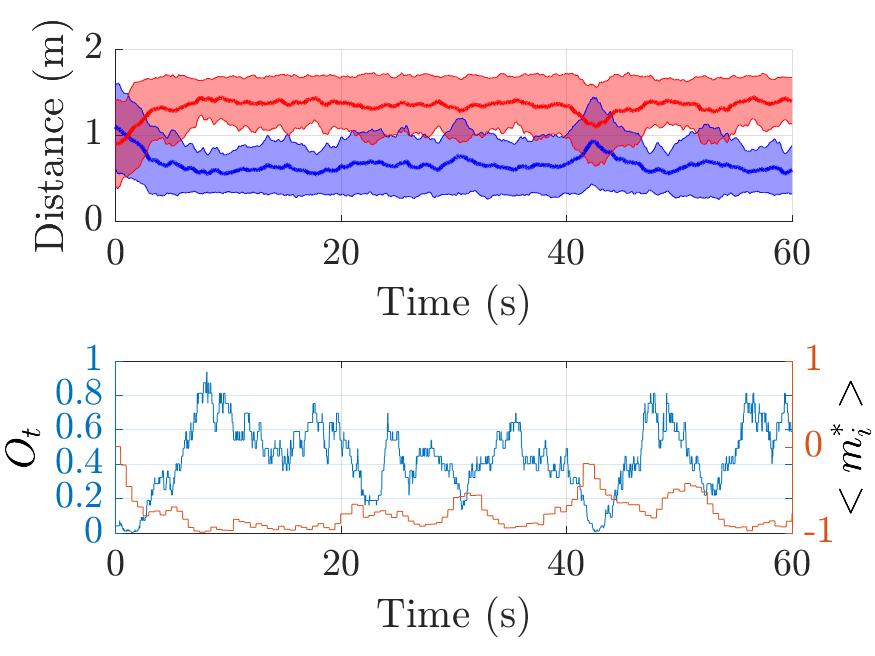}}
\resizebox{\columnwidth}{!}{\includegraphics{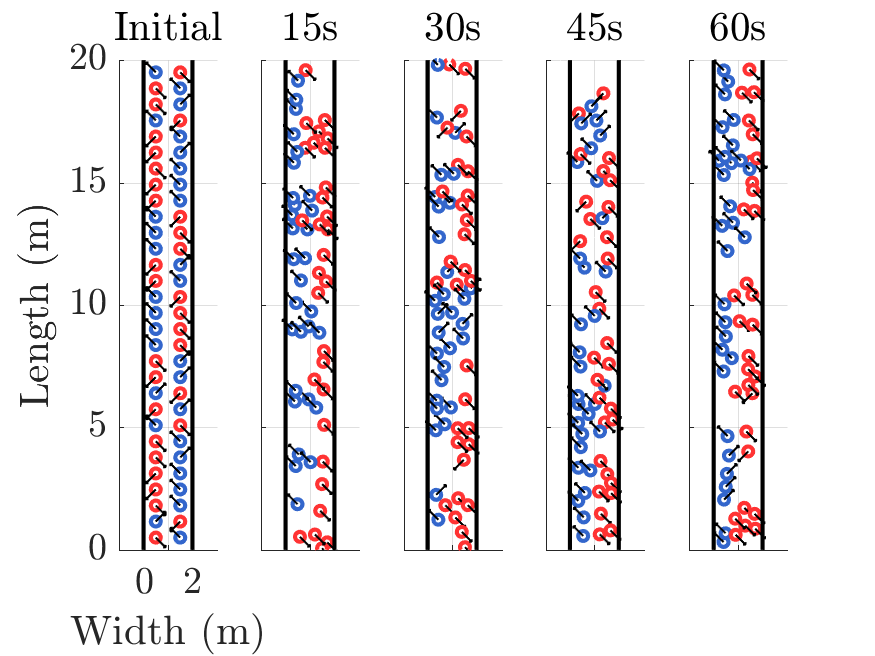}}
\caption{\label{Ordered} An example of a simulation where $\beta J = 2$, $h=0$ and $\sigma_p = 0.26$ ms$^{-1}$. (Top) The mean of position (and standard deviation) over time of the north (blue) and south (red) travelling pedestrians across the width of the corridor.  (Middle) The  order parameter $O_t$ as defined in equation~(\ref{OrderParameter}) and the mean expectation $\langle m_{i}^{*} \rangle$ of the pedestrians over time. (Bottom) A series of snapshots showing the configuration of pedestrians at 15s intervals. The arrows indicate the passing preferences of each pedestrian.}
\end{figure}

Figure~\ref{Ordered} summarises the outcome of a simulation with $h=0$, $\beta J = 2$ and $\sigma_p = 0.26$ ms$^{-1}$. This corresponds to a `polite' crowd where pedestrians put more weight on the observed preferences of others than their own. The figure shows the mean distance from the eastern wall for north (blue) and south (red) travelling pedestrians (and the corresponding standard deviation) as well as the order parameter $O_t$ as defined in equation~(\ref{OrderParameter}) and the mean of the individual expected preferences $\langle m_{i}^{*} \rangle$ at time $t$. In this particular simulation a preference for passing on the left hand side emerges within the first 10s. This coincides with an ordered state appearing, where those travelling north walk on the western side of the corridor (corresponding to their left) and those travelling south walk on the eastern side (corresponding to their left) in two lanes. This state is sufficiently stable that it prevents any further mixing between the two lanes. In contrast, the results of a simulation run with the social force model and the same heterogeneity in desired walking speed ($\sigma_p = 0.26$ ms$^{-1}$), but {\em without} the passing model does not order. It is evident then that the passing model (\ref{ising}) is generating the symmetry breaking observed in Figure~\ref{Ordered}.

\begin{figure}[h]
\resizebox{\columnwidth}{!}{\includegraphics{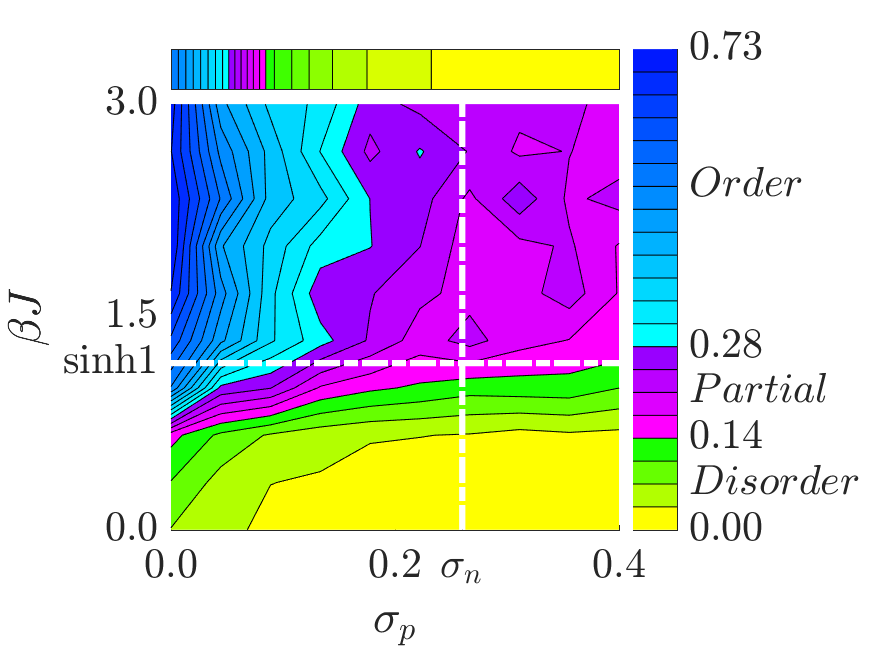}}
\caption{\label{PhaseSymmetric}The phase diagrams as function of $\beta J$ and $\sigma_p$ for: (bottom) the social force model with binary decision-making and passing, but with $h=0$; (top) the social force model \cite{castellano2009statistical} without passing. The different states of ordering (ordered, partially ordered and disordered) have been highlighted.}
\end{figure}

By varying $\beta J$ and $\sigma_p$ in our simulations with the passing model (while $h=0$) we construct a phase diagram for the system as shown in Fig.~\ref{PhaseSymmetric}. The diagram reveals three `phases': ordered (as seen in the example in Fig.~\ref{Ordered}), partially ordered and disordered. We observe two phase transitions between these states, one induced by the change in $\sigma$, and the other induced by the change in $\beta J$. The two dashed lines on the phase diagram indicate the location of phase transition as anticipated by equation~(\ref{weiss}) (at $\beta J = \sinh(1)$) and the experimentally observed heterogeneity of desired walking speeds $\sigma_{p} = 0.26$ ms$^{-1}$ \cite{henderson1971statistics}. Note that the transition from order to disorder that is driven by an increase in the heterogeneity in walking speeds also occurs in simulations without the decision-making and passing model (top, Fig.~\ref{PhaseSymmetric}) but at much lower heterogeneities ($\sigma_p \sim 0.10$ ms$^{-1}$). The effect of the decision-making and passing model is to stabilise the formation of columns in pedestrian flow.  

\begin{figure}[h]
	\resizebox{\columnwidth}{!}{\includegraphics{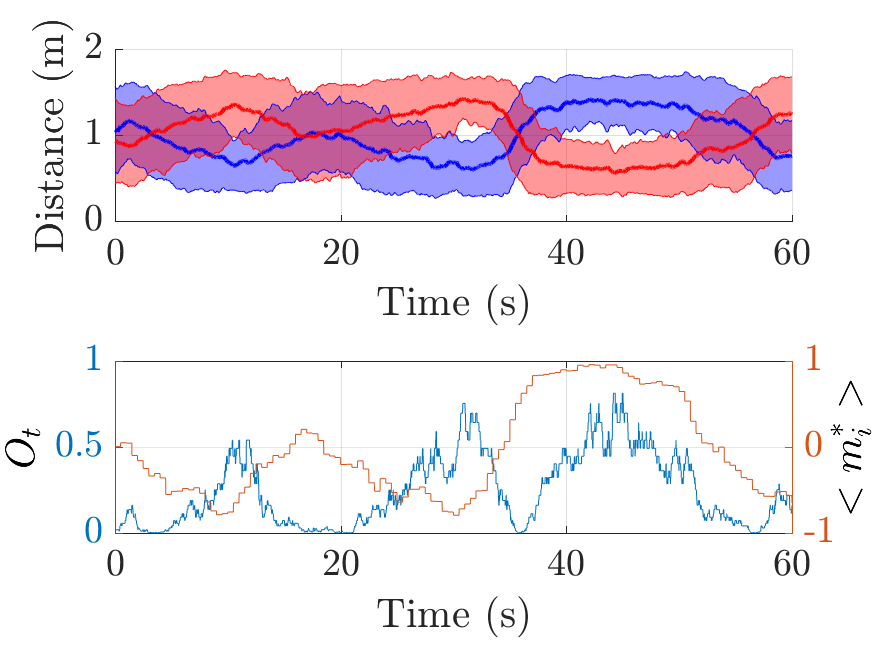}}
	\resizebox{\columnwidth}{!}{\includegraphics{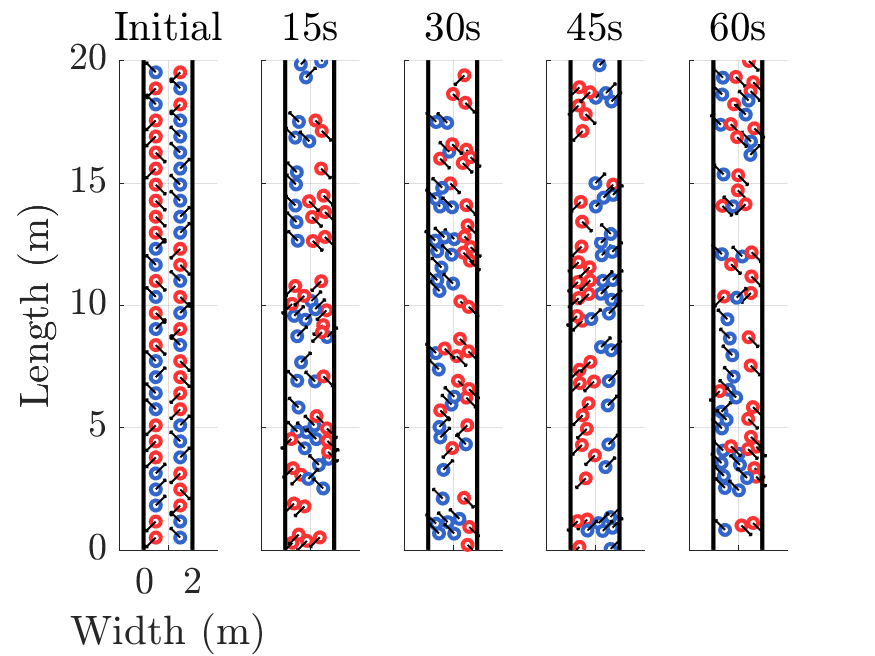}}
\caption{\label{PartialOrdered} Example of a partially ordered system where $\beta J = \sinh(1)$, $h=0$ and $\sigma = 0.26$ ms$^{-1}$. We observe the switching of lanes as clusters of pedestrians switch their preferences. }
\end{figure}

The partially ordered state is characterised by flips in pedestrian passing preference as shown in Fig.~\ref{PartialOrdered}. In a partially ordered system, $\beta J$ is high enough to overcome a pedestrian $i$'s own preferred passing direction, as long as $i$ does not encounter a group of pedestrians attempting to pass with different passing directions where $m_{i}^\star \sim 0$. Partial ordering is induced by a high spread in desired walking speeds and a value of $\beta J \geq \sinh(1)$ (as seen in Fig.~\ref{PhaseSymmetric}). A consensus in preferred passing direction can emerge, but can be broken when pedestrians become isolated, reverting to their own preference as $m_{i}^\star \rightarrow 0$. This can then induce mixing between the two lanes, reducing the overall order of the system, and leading the direction of the two lanes to flip. When $\beta J \ll 1$ the pedestrians are inattentive to the preferences of others and a preferred passing direction does not emerge in the simulations (as expected from equation~\ref{weiss}). This is the disordered phase.

\begin{figure}[h]
	\resizebox{\columnwidth}{!}{\includegraphics{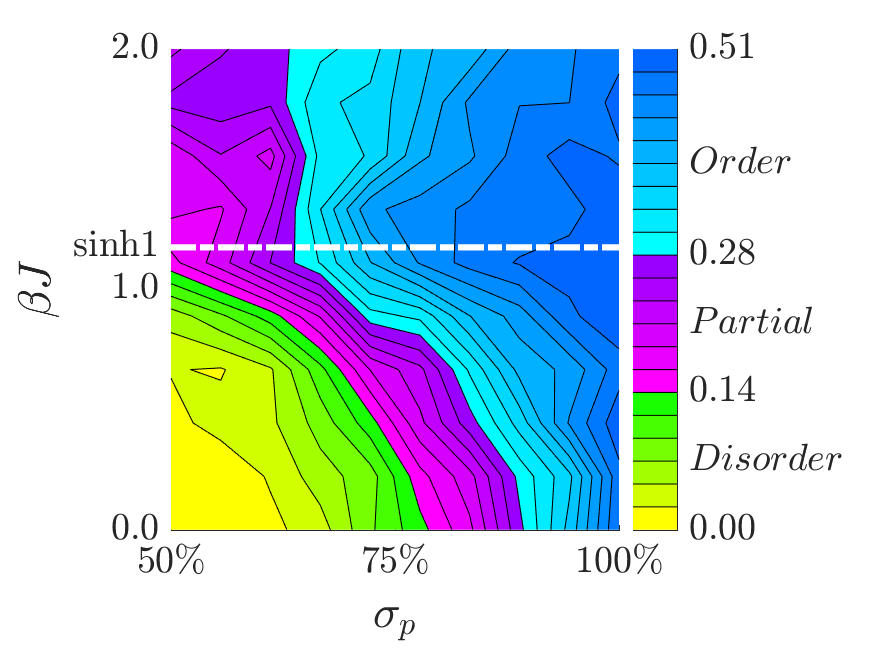}}
	\caption{\label{Skew} Phase diagram as a function of a skew in intrinsic passing preferences ($h \geq 0$), expressed as the percentage of pedestrians that favour the left, and $\beta J$ for $\sigma_p = 0.26$ ms$^{-1}$. At 50\% skew, $h = 0$ (i.e. an even split of pedestrians with intrinsic preferences for the left and right) and we recover the results shown in Fig.~\ref{PhaseSymmetric}. }
\end{figure}

Finally, we consider the influence of adding a bias $h_i = h$ to the system as per equation~(\ref{ising}). This is equivalent to simulating a pedestrian population with a skewed distribution of intrinsic passing preferences $\epsilon_i$, so we describe the bias in terms of the proportion of pedestrians who effectively prefer the left to the right. Fixing $\sigma_p = 0.26$ ms$^{-1}$ results in a second phase diagram as a function of $\beta J$ and the proportion of pedestrians favoring the left as shown in Fig.~\ref{Skew}. Again we observe the three states (order, partial order, and disorder) that were also seen in Fig.~\ref{PhaseSymmetric}. As might be expected, the skew in preferences increases the stability of the lanes, impeding flips in preferences. All three phases exist up to a skew of 80\% towards the left, at which point the completely disordered phase disappears for all values of $\beta J$. At a skew of 90\% only the completely ordered state remains. We note that this ordered state persists even when the variability of walking velocities is high. 

These results show that decisions about whether to pass other pedestrians on the left or the right may have a significant effect on the formation of lanes in pedestrian flows. We have assumed that individual pedestrians have developed their own preferences for passing one way or the other, whether through a longer-term process of social norm setting or otherwise. Over time, one might expect individual preferences to habituate to a group consensus, stabilising lane formation and skewing the breaking of symmetry. However, even when there is a high degree of heterogeneity in individual preferences, if pedestrians are attentive to the choices of others ($\beta J \gg 1$), a consensus can still emerge that stabilises lane formation. We also observed switching of formed lanes from left to right and vice versa, a phenomenon observed in experiments \cite{moussaid2012traffic}. In our simulations, the switching of lanes was accompanied by a temporary switch in preference.

To summarise, we have used a simple binary decision-making model to investigate the emergence of preferred passing directions in pedestrian flows. With perfectly polarised intrinsic preferences, we show that a preferred passing direction can emerge provided pedestrians are sufficiently attentive to the preferences of others. We have incorporated this model into a social force simulation of pedestrian dynamics, and shown that this has a significant impact on the formation of ordered patterns of flow. If a consensus emerges, traffic in a narrow corridor settles into co-moving lanes, otherwise symmetry is preserved and the system is disordered, resulting in less efficient transport. We have also identified a third phase, where the preferred passing direction flips between left and right, so that the symmetry is broken instantaneously but not permanently.  


\bibliography{Pedestrians}
   
\end{document}